# Zonas costeras bajas en el Golfo de México ante el incremento del nivel del mar

*J.A. Carbajal Domínguez*

## Resumen

En este trabajo se presenta una estimación de la extensión territorial afectada en el caso de incrementos del nivel del mar de 0.6 m, 1 m y 2 m. Asimismo, se estima el número de habitantes en las zonas vulnerables que se producirían en el escenario de elevación del mar en 1 m. Para ello, se utilizan los datos de elevación del terreno de la Shuttle Radar Topography Mission de la Nasa junto con un algoritmo propio que permite la reconstrucción de la zona afectada. Para la estimación de la superficie se emplea el procesamiento digital de imágenes para la delimitación de dichas zonas. Los resultados se procesan geo-referenciados para compararlos con los asentamientos humanos en la región de interés. Los resultados muestran que el área afectada total corresponde al 1.26% del a extensión territorial nacional y al 3.18% de la población total del país. Se verán afectadas 174 poblaciones de más de 1 000 habitantes. El estado con mayor superficie afectada es Tabasco con más del 21% de su territorio, mientras que el quien tendrá una población más vulnerable será Veracruz, con más de 1 millón de habitantes si el escenario de incremento en 1m se presentará hoy en día. Quintana Roo, por su parte, tendrá proporcionalmente un mayor impacto pues sufrirá los efectos en el 81.1% de su población. Se listan las poblaciones vulnerables y se muestran los mapas correspondientes a las diferentes zonas estudiadas.

**Palabras clave**: zona costera, incremento nivel dem mar, Golfo de México

## Objetivo

Determinar la extensión territorial y la población afectada ante el incremento del nivel del mar en escenarios de 0.6 m, 1 m y 2 m en la costa del Golfo de México y el Caribe

## Objetivos específicos

- Obtener y procesar información de elevación de terreno para la costa los estados de Tamaulipas, Veracruz, Tabasco, Campeche, Yucatán y Quintana Roo.
- Desarrollar un método de estimación de la extensión territorial vulnerable en los escenarios considerados.
- Determinar la extensión territorial vulnerable para cada estado considerado.
- Determinar el número de habitantes y las poblaciones más vulnerables ante el escenario de elevación del nivel del mar de 1m.

## Metodología

### Materiales

Para el presente estudio se emplean los archivos de datos con formato hgt para disponibles en el sitio de usgs (http://dds.cr.usgs.gov/srtm/version2_1/srtm3/North_America/), relativos a los estados de la costa del Golfo de México y del Caribe mexicano, la lista de todos los archivos utilizados se muestra en el apéndice I. Cabe mencionar que estos archivos contienen los datos de información de elevación de terreno obtenidos en la misión del transbordador espacial (Shuttle Radar Topography Mission, http://www2.jpl.nasa.gov/srtm/p_status.htm). Por otro lado, los datos de población se obtienen de Inegi (2005)[1] para los estados de Tamaulipas, Veracruz, Tabasco, Campeche, Yucatán y Quintana Roo.

Se descargan de la página del Conacyt Queretaro (http://www.concyteq.edu.mx/cqrn2/kmldownload.htm) los archivos kml con los polígonos que delimitan la extensión territorial de los estados y municipios de los estados del Golfo.

El procesamiento se realiza en ambiente Windows, con programas en c para el procesamiento de los datos de terreno y en lenguaje python para la escritura y despliegue de la información en archivos tipo kml (KeyHole Markup Language) para realizar la geo-referenciación de los resultados obtenidos.

Para el procesamiento y los cálculos se emplea una computadora pc con procesador amd phenom de triple núcleo y 2 gb de memoria ram, con disco duro de 300 gb.

---

[1] Conteo de Población Inegi 2005. Principales resultados por localidad 2005 (iter), http://www.inegi.org.mx/est/contenidos/espanol/sistemas/conteo2005/localidad/iter/default.asp?s=est&c=10395.

Para visualizar los resultados, se emplea el programa World Wind ( http://worldwind.arc.nasa.gov/java/para la visualización espacial de los resultados.

## Descripción de la metodología utilizada

Como ya se mencionó previamente, los archivos DEM (Digital Elevation Models), utilizados son los SRTM (Shuttle Radar Topography Mission) de la Nasa y que están disponibles gratuitamente via ftp en USGS (http://dds.cr.usgs.gov/srtm/version2_1/srtm3/North_America/). Estos archivos comprenden 1° lat. por 1° long. Su nomenclatura se refiere a la esquina sur-oeste de este (esquina inferior izquierda) de este cuadrado. Cada uno de ellos contiene del orden de 1,214 x 1,214 mediciones, aunque estos valores cambian en cada caso, por lo que se tiene aproximadamente una medida elevación cada 100 m. Cada uno de estos archivos es transformado con un programa propio en dos archivos de texto: uno con la información de la información de tamaño de archivo y coordenadas, y otro con los datos de elevación del terreno en coordenadas UTM.

Los datos de elevación son procesados para determinar las zonas menores o iguales a la cota de incremento del nivel del mar considerada.

Sin embargo, de esta forma se obtiene sólo un muestreo discreto de datos o puntos de una superficie bidimensional. Esto hace necesario construir un algoritmo propio que permita delimitar las áreas de afectación definidas por estos puntos debido a que no se encontró reportado ninguno en la literatura. Esto a pesar de que un estudio similar de áreas afectadas ante el incremento del nivel del mar se encuentra disponible en el sitio del Departement of Geosciences Enviromental Studies Laboratory (DGESL) de la Universidad de Arizona (Departament of Geosciences Enviromental Studies Laboratory, http://geongrid.geo.arizona.edu/arcims/website/slrworld/viewer.htm).

El algoritmo propuesto aquí consiste en convolucionar los datos que cumplan el criterio de la cota de elevación especificada con una función gaussiana de radio R, como se indica en las ecuaciones (1) y (2) (Goodman, 2004)[2],

$$A(x,y) = C(x,y) \otimes G(x,y) \qquad (1)$$

en donde $A(x,y)$ el área afectada, $C(x,y)$ son los datos que cumplen con el criterio de la elevación requerida, $\otimes$ denota el producto de convolución y $G(x,y)$ es una función gaussiana en 2D dada, en este caso, por

$$G(x,y) = exp\left[-(x^2 + y^2)/R^2\right] \qquad (2)$$

Con $R$ un parámetro que define el radio de extensión de la afectación alrededor de un punto de medición. En este caso se considera $R=10$ equivalente aproximadamente a 1000 m. La convolución *(1)* se realiza multiplicando escalarmente las transformadas de Fourier de cada una de las funciones descritas y finalmente tomando la transformada de Fourier inversa de este producto. Además, como lo que interesa es la extensión y no la amplitud de la función *A(x,y)* se escala toda la intensidad a 1. Un ejemplo de procesamiento se presenta en la figura 1. Con los resultados se escribe un archivo de imagen en formato jpg, para tener una mayor eficiencia.

---

[2] Goodman, JW., 2004. Introduction to Fourier Optics. Roberts & Company Publishers.

Como se observa en la figura 1, los puntos de elevación que satisfacen el criterio de elevación se encuentran distribuidos como se muestra en (A), mientras que después de aplicar la convolución, se tiene una superficie bien definida, como se muestra en (B). Además, se puede observar que se forman áreas conexas y otras que aparecen aisladas, como las dos zonas que se aprecian en la parte inferior izquierda. Eso indica que la zona principal de afectación se puede extender hasta esos lugares.

Ya con los resultados, se genera un archivo en lenguaje kml que permite su despliegue en un sistema de información geográfica utilizando la imagen previamente obtenida, en este caso el software World Wind aunque el formato kml también puede ser utilizado por programas (*e.g.* Google Earth). En este mismo formato se emplean los datos de los polígonos que definen los estados y municipios de interés para el presente estudio. De esta forma, al superponer los resultados de las zonas vulnerables junto con los datos de los límites territoriales es posible determinar las áreas vulnerables por estado y por municipio.

Para determinar el área afectada por estado, se considera la imagen del estado total y se binariza (uno para puntos en el estado y cero para los puntos fuera) (González y Woods, 2007)[3] junto con la imagen de las áreas vulnerables del estado. El porcentaje del área afectada puede estimarse mediante el cociente de las sumas de los píxeles,

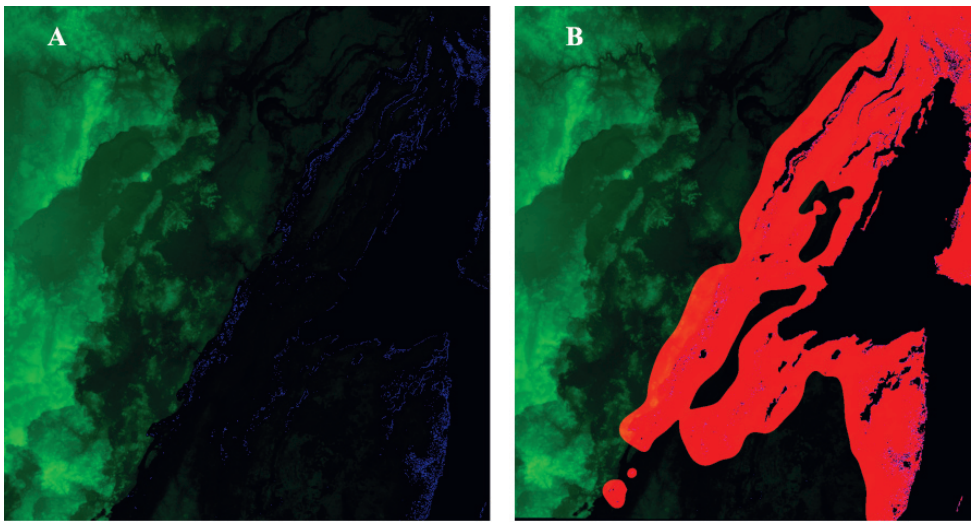

**Figura 1.** Imagen generada con datos de elevación de terreno de la bahía de Chetumal. A) En verde se muestra el terreno; las zonas más brillantes corresponden a zonas más altas. En azul se muestran los puntos cuya altura es menor o igual a 0.6m. Se observa que siguen una distribución discreta. B) Resultado de aplicar el algoritmo: en rojo se muestra la posible extensión del terreno cuya elevación cumpla con los criterios deseados. Se observa una definición continua. En este ejemplo R=10 pixeles que corresponden a 1000m aproximadamente.

$$A\% = \left(\frac{\Sigma Pv}{\Sigma PT}\right) \times 100\% \quad (3)$$

En la que *A%* es el porcentaje del área afectada, *Pv* son los píxeles que conforman el área vulnerable y *PT* los píxeles que conforman el área del estado.

Para verificar estos resultados, se realiza un procedimiento similar para imágenes de la Universidad de Arizona[4] para los estados de interés y para elevaciones del mar de 1 y 2 m aunque en este caso se realizo además un procesamiento en color para aislar las zonas vulnerables. Esto es, se convierte la imagen a color en RGB y se transforma al espacio HSV en el que se aíslan los píxeles correspondientes al color rojo. Un ejemplo se muestra en la figura 2.

Para determinar el número de habitantes vulnerables, se utilizan los datos del Inegi del conteo de población 2005, para los estados de Tamaulipas, Veracruz, Tabasco, Campeche, Yucatán y Quintana Roo. Como primer paso, se consideron únicamente los datos de los municipios cuyo polígono coincide o contiene parte del área de afectación calculada para el caso de 1 m debido a que es el escenario más probable en el mediano plazo. De cada municipio se tomaron en consideración —en aras de facilitar el cálculo— la información correspondiente a las poblaciones mayores a 1 000 habitantes así como sus respectivas coordenadas geográficas. Con un programa propio, los datos son empleados para escribir una archivo con la información de las coordenadas en formato kml para poder desplegarles en un mapa.

Posteriormente, se descarta a las poblaciones que no están dentro o que no estén suficientemente cerca del área de afectación. Aquí se entiende que los asentamientos no son puntuales, si no que tienen una importante extensión llamada comúnmente mancha urbana. Por lo tanto, se consideran puntos cuya mancha urbana quede en la vecindad de las zonas de afectación. De esta forma se obtiene una lista de las poblaciones afectadas en el caso de tal escenario.

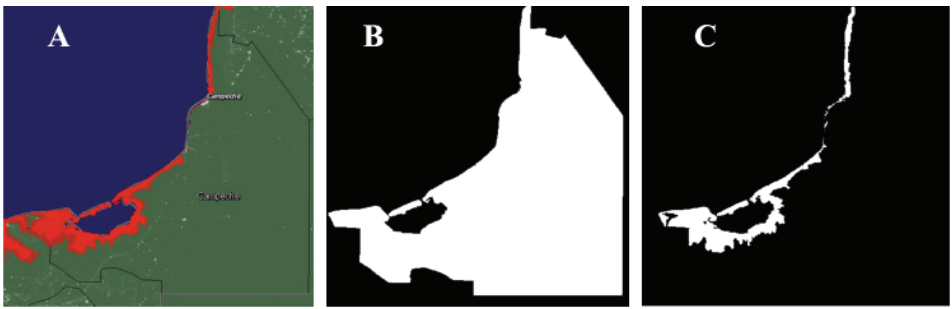

**Figura 2.** Ejemplo del procesamiento. A) Imagen del estado que se desea analizar; B) imagen binarizada del estado. Aquí el número de pixeles blancos es proporcional a la superficie del estado; C) Resultado del procesamiento de color para dejar únicamente los pixeles rojos. El número de estos pixeles es proporcional al área de afectación.

---

[3] González, RC., y R.E. Woods, 2007. Digital Image Processing. Prentice Hall.
[4] Departament of Geosciences Enviromental Studies Laboratory, http://geongrid.geo.arizona.edu/arcims/website/slrworld/viewer.htm

# Resultados y discusión

## Superficie afectada

Los resultados obtenidos para cada estado, en cada escenario, se muestran en la tabla 1, la cual contiene además a manera de comparación, los resultados que se obtuvieron procesando las imágenes de los mapas de la Universidad de Arizona para las mismas zonas.

Se observa que ambos métodos coinciden satisfactoriamente para Quintana Roo, Yucatán, Veracruz y Tamaulipas. Sin embargo, las diferencias son notables para los casos de Tabasco y Campeche. Esto se debe a que en la zona fronteriza entres ambos estados, los datos de elevación de terreno tienen errores debido a que existe una importante cobertura vegetal, así como en las regiones bajas que incluso se encuentran por debajo del nivel del mar, las cuales permanecen cubiertas de agua la mayor parte del año. Debido a que los datos fueron obtenidos mediante interferometría de radar, la vegetación y los cuerpos de agua tienden a modificar los resultados. El efecto es que en el método de la Universidad de Arizona, la región afectada parece subestimada, pues se excluye a las regiones de Jonuta, Macuspana y Emiliano Zapata pertenecientes al estado de Tabasco, así como el área correspondiente a Palizada en Campeche. Mientras que en el proceso aquí reportado dichas áreas son incluidas.

Para los otros estados, los dos métodos coinciden satisfactoriamente pues la diferencia entre la estimación de ambos métodos es menor al 3%. Lo anterior demuestra que el método aquí presentado para estimar el área afectada a partir de los datos de elevación es bastante razonable. La tabla 1 muestra el concentrado de los resultados en orden descendente de las áreas afectadas para los escenarios de 0.6 m, 1 m y 2 m.

En ella se muestra que, el estado de Tabasco es el que sufre una mayor área afectada, la cual se sitúa entre el 20 y el 25% del área total del estado. Mientras que Campeche y Quintana Roo tienen una disminución de alrededor 12% de su superficie. Finalmente, Yucatán, Veracruz y Tamaulipas muestran una menor afectación, al situarse entre 2 y 6%.

Para el escenario de 1 m, se tiene que la superficie total afectada es igual al 1.26% de la extensión continental de los Estados Unidos Mexicanos.

| Tabla 1. Porcentaje de la superficie afectada para cada estado en los distintos escenarios. | | | | | |
|---|---|---|---|---|---|
| **Estado** | **0.6m** | **1m** | **2m** | **1m*** | **2m*** |
| Tabasco | 21.44% | 25.87% | 25.45% | 8.18% | 14% |
| Campeche | 12.46% | 12.60% | 15.00% | 7.46% | 9.50% |
| Quintana Roo | 11.38% | 12.05% | 12.50% | 9.47% | 11.94% |
| Yucatán | 3.53% | 3.86% | 5.88% | 4.70% | 6.27% |
| Veracruz | 3.21% | 3.50% | 4.64% | 5% | 5.30% |
| Tamaulipas | 2.29% | 2.40% | 3.29% | 2% | 3.60% |
| *resultados de superficie con imágenes de la Universidad de Arizona | | | | | |

## Población afectada

En la tabla 2 se concentran los resultados obtenidos para el número de habitantes afectados en el escenario de 1m. Se muestra que Veracruz tendrá la afectación de un mayor número de habitantes que los otros estados. Le siguen Tamaulipas y Quintana Roo. La tabla 2 también muestra la densidad de habitantes vulnerables, la cual es la razón entre el número de habitantes de las zonas vulnerables y la extensión en km² de dicha área. Este parámetro permite ver cuáles estados tendrán una mayor presión sobre asentamientos humanos densamente poblados y que seguramente deberán implementar estrategias de reubicación. En ambos casos, aunque su superficie territorial sufre una afectación relativamente pequeña, dicha área cuenta con asentamientos humanos importantes.

En cuanto al porcentaje de la población vulnerable de cada estado, se tiene que Quintana Roo y Campeche verán afectados el 81.10% y 58.41% de su población respectivamente. Esto se debe a que la mayor parte de su población se asienta en ciudades cercanas a la costa.

En el caso de Tabasco y Yucatán, se observa que son los que tienen una menor población afectada, debido probablemente, a que las características de las áreas vulnerables ante el incremento del nivel del mar, no han sido aptas históricamente para el asentamiento de grandes grupos poblacionales, debido a que, o son zonas permanentemente inundadas, como en el caso de Tabasco; o son zonas permanentemente expuestas a eventos hidrometeorológicos tales como los huracanes, en el caso de Yucatán, entre otros factores.

En la columna del número de asentamientos, se puede observar que el estado con un mayor número de localidades –mayores a 1 000 habitantes- afectadas es Tabasco con 61, seguido por Veracruz con 45. Si se considera que Tabasco tiene una franja costera mucho menor que Veracruz, se puede notar la alta dispersión poblacional de Tabasco.

La población total afectada de la costa del Golfo de México corresponde al 3.18% de la población total del país, asentadas en 174 localidades que deberán tomar medidas para protegerse de los riesgos inherentes a un escenario de incremento del nivel del mar.

| | Estado | Habitantes vulnerables | % población | No. Asentamientos | Densidad Habitantes/km² |
|---|---|---|---|---|---|
| | **Tabla 2**. Resultados de la población afectada por estado. | | | | |
| 1 | Veracruz | 1 008 928 | 14.18% | 45 | 401.37 |
| 2 | Tamaulipas | 651 647 | 21.54% | 13 | 338.66 |
| 3 | Quintana Roo | 920 772 | 81.10% | 23 | 180.38 |
| 4 | Campeche | 440 910 | 58.41% | 19 | 60.41 |
| 5 | Yucatán | 66 763 | 3.67% | 13 | 43.66 |
| 6 | Tabasco | 199 491 | 10.02% | 61 | 31.17 |
| | **Total** | 3 288 511 | | **174** | |
| | **% Pob. Nacional** | **3.18%** | | | |

A continuación, de las tablas 3 a la 8, se muestran los asentamientos humanos vulnerables por estado con su respectivo número de habitantes, se cuerdo al conteo de población Inegi (2005). Se incluyen las coordenadas geográficas de las mismas para evitar los errores por homonímias.

De las figuras 3 a la 15, se muestran los mapas de la zona estudiada con las poblaciones ahí asentadas. Las zonas vulnerables se muestran en rojo, mientras que en blanco se muestran las fronteras de los estados y los municipios. Los círculos pequeños muestran la ubicación geográfica de las poblaciones, de acuerdo a sus coordenadas. Algunas de las poblaciones que no aparecen claramente dentro de la zona de afectación se incluyen en este estudio debido a la cercanía a la zona de su mancha urbana.

## Conclusiones

Se ha obtenido una estimación para la superficie afectada en la costa del Golfo de México para los escenarios de incremento del nivel del mar de 0.6 m, 1 m y 2 m usando un método basado en el producto de convolución. Así mismo, se obtuvo una estimación de la población total vulnerable de esta región para el escenario de 1 m. Los resultados muestran que el área afectada total corresponde al 1.26% de la extensión territorial nacional y al 3.18% de la población total del país. Se verán afectadas 174 poblaciones de más de 1 000 habitan-

| | **Tabla 3.** Resultados de población afectada por localidad para Tamaulipas. | | | |
|---|---|---|---|---|
| | **Nombre** | Habitantes | Latitud | Longitud |
| 1 | Tampico | 303 635 | 22.2552778 | -97.8686111 |
| 2 | Ciudad Madero | 193 045 | 22.2763889 | -97.8313889 |
| 3 | Miramar | 82 079 | 22.3375 | -97.8694444 |
| 4 | Altamira | 50 896 | 22.3958333 | -97.9369444 |
| 5 | Soto la Marina | 9 389 | 23.7675 | -98.2077778 |
| 6 | Carboneras (La Carbonera) | 2 723 | 24.6263889 | -97.7166667 |
| 7 | Las Higuerillas | 2 036 | 25.2622222 | -97.4361111 |
| 8 | La Pesca | 1 632 | 23.7872222 | -97.7766667 |
| 9 | La Colonia (Estación Colonias) | 1 435 | 22.4358333 | -98.0172222 |
| 10 | Ricardo Flores Magón | 1 383 | 22.4530556 | -97.9055556 |
| 11 | Lomas del Real | 1 216 | 22.5194444 | -97.8994444 |
| 12 | San Germán | 1 153 | 25.2161111 | -97.9208333 |
| 13 | Carvajal | 1 025 | 24.5041667 | -97.7430556 |
| | **Total** | 65 1647 | | |
| | **%Pob.** | 21.54% | | |

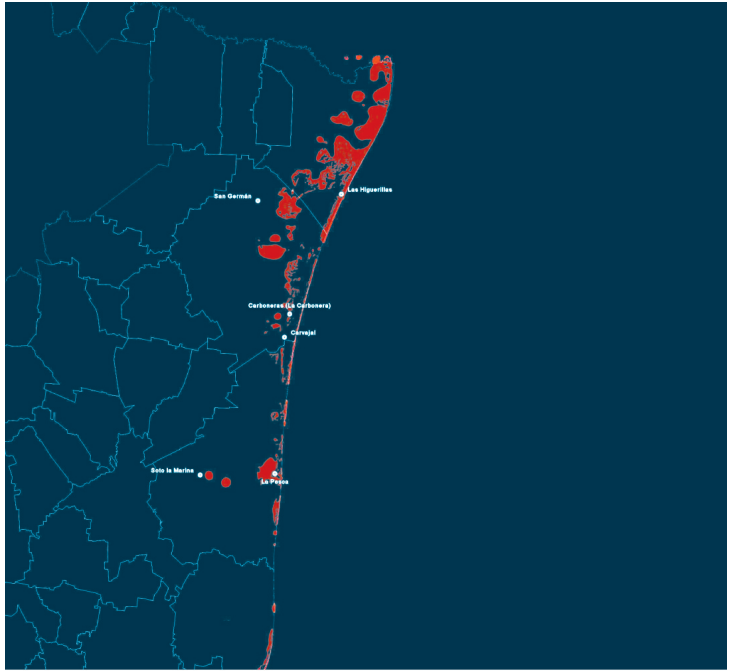

**Figura 3.** Tamaulipas, región norte. Las Higuerillas, Carboneras, Carvajal y La Pesca aparecen claramente en la zona de afectación.

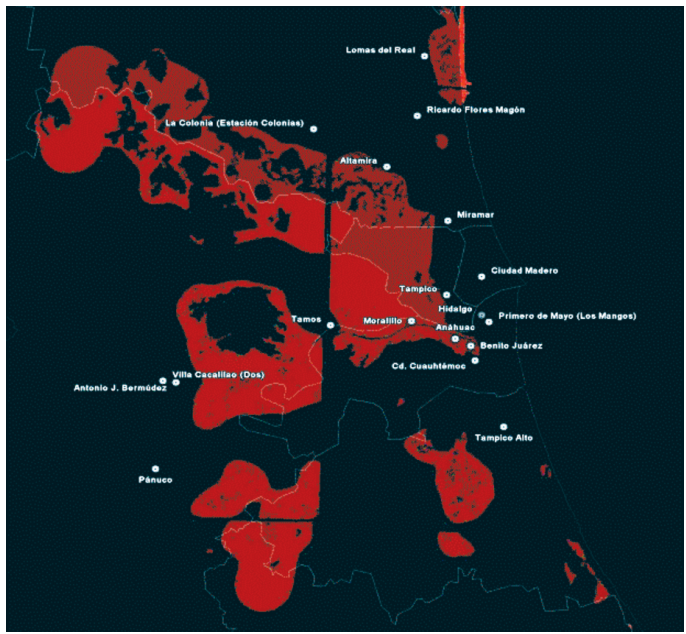

**Figura 4.** Frontera Tamaulipas-Veracruz. Esta zona constituye un área densamente poblada. En particular, el municipio de Tampico aparece ampliamente afectado.

**Tabla 4.** Resultados de población afectada por localidad para Veracruz.

|   | Nombre | Habitantes | Latitud | Longitud |
|---|---|---|---|---|
| 1 | Veracruz | 444 438 | 19.1991667 | -96.1377778 |
| 2 | Coatzacoalcos | 234 174 | 18.1377778 | -94.4352778 |
| 3 | Túxpam de Rodríguez Cano | 78 523 | 20.9588889 | -97.4030556 |
| 4 | Pánuco | 37 450 | 22.0552778 | -98.1775 |
| 5 | Alvarado | 22 330 | 18.7702778 | -95.7605556 |
| 6 | Allende | 20 501 | 18.15 | -94.405 |
| 7 | Benito Juárez | 14 015 | 22.1994444 | -97.8411111 |
| 8 | Anáhuac | 13 657 | 22.2069444 | -97.8580556 |
| 9 | Alto Lucero | 13 525 | 20.9508333 | -97.4433333 |
| 10 | Gutiérrez Zamora | 13 484 | 20.4491667 | -97.0838889 |
| 11 | Boca del Río | 10 980 | 19.1008333 | -96.1072222 |
| 12 | Moralillo | 9 154 | 22.2255556 | -97.9058333 |
| 13 | Cd. Cuauhtémoc | 8 950 | 22.1833333 | -97.8361111 |
| 14 | Santiago de la Peña | 8 538 | 20.945 | -97.4063889 |
| 15 | Tlacotalpan | 8 006 | 18.6116667 | -95.6611111 |
| 16 | Antón Lizardo | 6 187 | 19.0566667 | -95.9880556 |
| 17 | Hidalgo | 6 159 | 22.2338889 | -97.8302778 |
| 18 | Primero de Mayo (Los Mangos) | 5 068 | 22.2263889 | -97.8222222 |
| 19 | Fraccionamiento Ciudad Olmeca | 4 948 | 18.1497222 | -94.5536111 |
| 20 | Tamiahua | 4 849 | 21.2780556 | -97.4455556 |
| 21 | Tecolutla | 4 523 | 20.4797222 | -97.01 |
| 22 | Tamos | 3 740 | 22.2188889 | -97.9930556 |
| 23 | Nautla | 3 118 | 20.2072222 | -96.7722222 |
| 24 | Palma Sola | 2 633 | 19.7705556 | -96.4313889 |
| 25 | Tampico Alto | 2 242 | 22.1105556 | -97.8033333 |
| 26 | Villa Cacalilao (Dos) | 2 132 | 22.1530556 | -98.1722222 |
| 27 | Casitas | 2 024 | 20.2538889 | -96.7991667 |
| 28 | Tonalá | 1 989 | 18.2075 | -94.1388889 |
| 29 | Paso Nacional | 1 830 | 18.7677778 | -95.7475 |
| 30 | Cucharas | 1 592 | 21.6158333 | -97.6577778 |
| 31 | Antonio J. Bermúdez | 1 506 | 22.1516667 | -98.1580556 |
| 32 | Estero de Milpas | 1 467 | 21.2538889 | -97.4497222 |
| 33 | Saladero | 1 379 | 21.4247222 | -97.5433333 |
| 34 | La Victoria (La Peñita) | 1 371 | 20.9361111 | -97.3788889 |
| 35 | Las Escolleras | 1 334 | 18.7755556 | -95.7472222 |
| 36 | La Guadalupe | 1 218 | 20.3736111 | -96.9216667 |

**Tabla 4 (continuación).** Resultados de población afectada por localidad para Veracruz.

|    | Nombre | Habitantes | Latitud | Longitud |
|----|--------|------------|---------|----------|
| 37 | Rancho Nuevo | 1 209 | 20.6711111 | -97.2063889 |
| 38 | Las Higueras | 1 194 | 20.0361111 | -96.6211111 |
| 39 | Mandinga y Matoza | 1 154 | 19.0486111 | -96.0730556 |
| 40 | Banderas | 1 135 | 20.9883333 | -97.3933333 |
| 41 | Barra de Cazones | 1 065 | 20.7208333 | -97.2033333 |
| 42 | El Farallón | 1 063 | 19.6375 | -96.4102778 |
| 43 | Playa de la Libertad | 1 053 | 19.0822222 | -96.0972222 |
| 44 | Playa de Chachalacas | 1 015 | 19.4175 | -96.3247222 |
| 45 | Las Barrillas | 1 006 | 18.1863889 | -94.5961111 |
|    | **Total** | 1 008 928 | | |
|    | **%Pob.** | 14.18% | | |

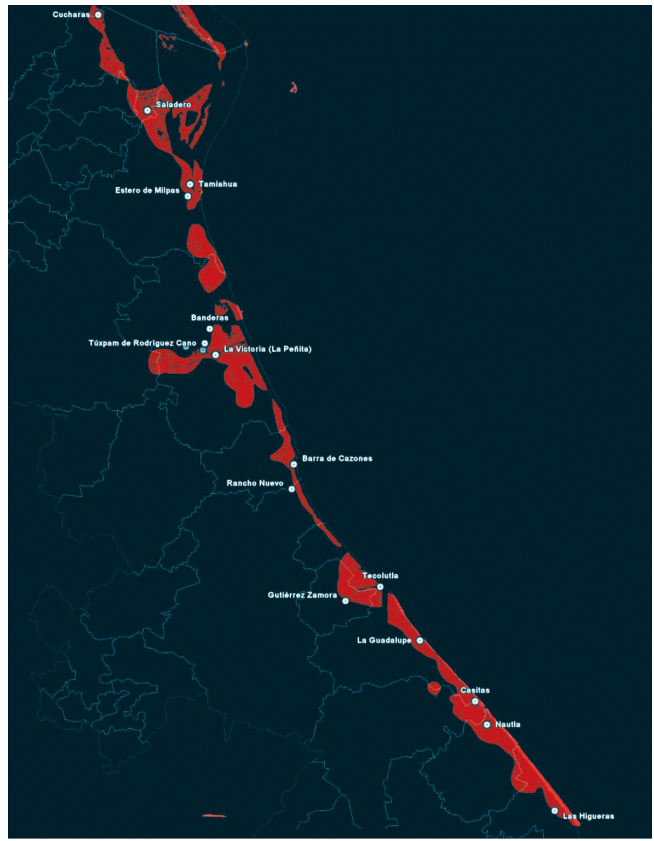

**Figura 5.** Veracruz. En sta región todas las poblaciones costeras aparecen en la zona afectada.

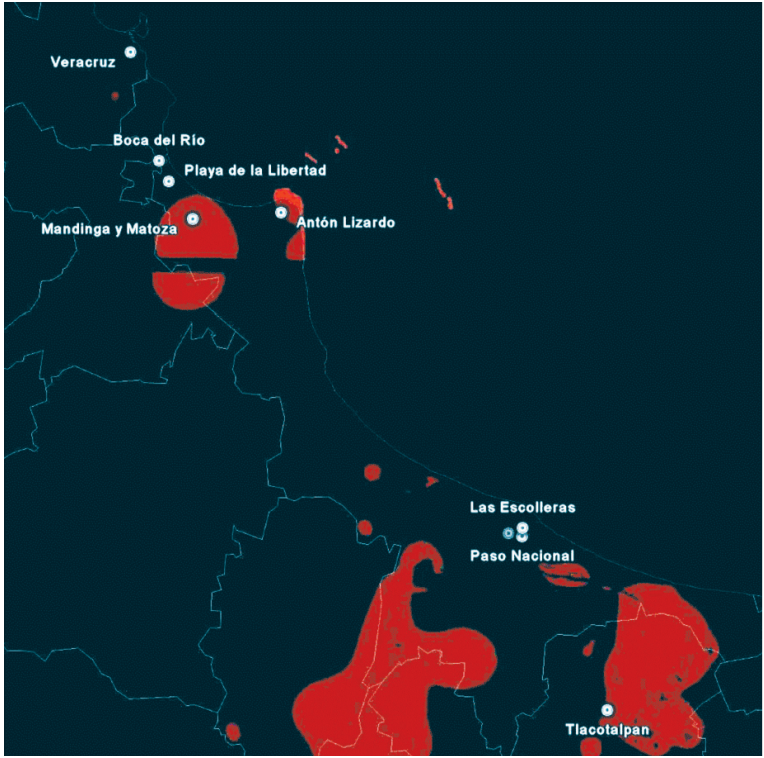

**Figura 6.** Mapa correspondiente a las regiones cercanas al puerto de Veracruz.

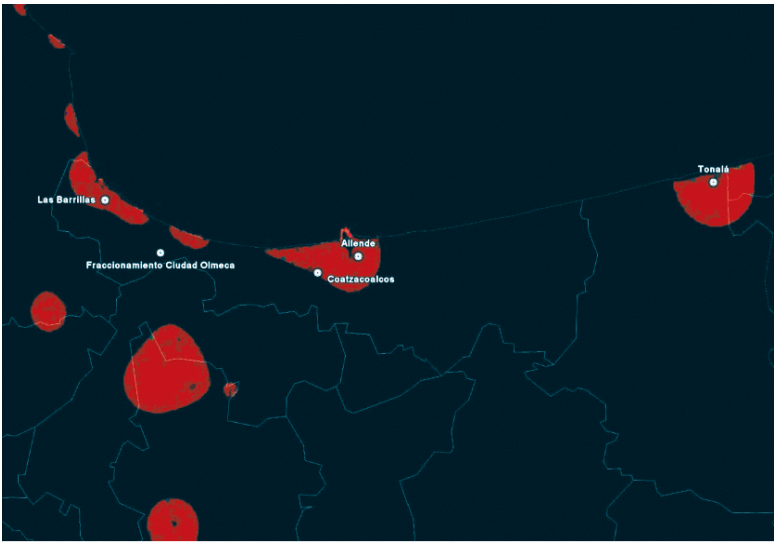

**Figura 7.** Veracruz-Tabasco. El municipio de Coatzacoalcos, densamente poblado, se muestra como una zona afectada.

**Tabla 5.** Resultados de población afectada por localidad para Tabasco.

| | Nombre | Habitantes | Latitud | Longitud |
|---|---|---|---|---|
| 1 | Paraíso | 24 773 | 18.3961111 | -93.2127778 |
| 2 | Frontera | 21 810 | 18.5336111 | -92.6469444 |
| 3 | Emiliano Zapata | 16 796 | 17.7413889 | -91.7636111 |
| 4 | Tamulté de las Sabanas | 7 874 | 18.1616667 | -92.7838889 |
| 5 | Vicente Guerrero | 7 554 | 18.3913889 | -92.8919444 |
| 6 | Cor. Andrés Sánchez Magallanes | 7 277 | 18.2933333 | -93.8633333 |
| 7 | Jonuta | 6 341 | 18.0902778 | -92.1366667 |
| 8 | Pemex (Ciudad Pemex) | 5 752 | 17.8822222 | -92.4825 |
| 9 | La Curva | 5 098 | 17.8686111 | -92.4883333 |
| 10 | Buena Vista 1a. Sección | 4 527 | 18.1441667 | -92.7497222 |
| 11 | Quintín Arauz | 4 341 | 18.3675 | -93.2147222 |
| 12 | Simón Sarlat | 4 329 | 18.3438889 | -92.8097222 |
| 13 | Reyes Hernández 2a. Sección | 3 506 | 18.2369444 | -93.2558333 |
| 14 | Cuauhtémoc | 3 405 | 18.4019444 | -92.955 |
| 15 | Ignacio Allende | 3 314 | 18.3830556 | -92.8444444 |
| 16 | Chablé | 3 152 | 17.8561111 | -91.7816667 |
| 17 | Guatacalca | 3 138 | 18.1669444 | -92.9780556 |
| 18 | Puerto Ceiba | 2 726 | 18.4113889 | -93.18 |
| 19 | Cocohital | 2 403 | 18.3972222 | -93.3455556 |
| 20 | Moctezuma 1a. Sección | 2 309 | 18.3766667 | -93.2313889 |
| 21 | Moctezuma 2a. Sección | 2 292 | 18.3686111 | -93.2208333 |
| 22 | Libertad 1a. Sección (El Chivero) | 2 246 | 18.3291667 | -93.1694444 |
| 23 | Ignacio Zaragoza | 2 090 | 18.4016667 | -92.9594444 |
| 24 | Álvaro Obregón (Santa Cruz) | 2 087 | 18.3886111 | -92.8022222 |
| 25 | Las Flores 1a. Sección | 1 838 | 18.4025 | -93.2291667 |
| 26 | Francisco I. Madero | 1 795 | 18.4663889 | -92.7416667 |
| 27 | Tucta | 1 790 | 18.195 | -92.9938889 |
| 28 | Nicolás Bravo | 1 789 | 18.2941667 | -93.13 |
| 29 | José María Pino Suárez 1a. Sección | 1 770 | 18.3522222 | -93.3822222 |
| 30 | Monte Grande | 1 757 | 17.9358333 | -92.2638889 |
| 31 | Las Flores 2a. Sección | 1 712 | 18.4244444 | -93.2519444 |
| 32 | Benito Juárez | 1 685 | 18.4225 | -92.8058333 |
| 33 | Zapotal 2a. Sección | 1 653 | 18.3130556 | -93.2666667 |
| 34 | Puerto Ceiba (Carrizal) | 1 648 | 18.405 | -93.1897222 |
| 35 | Olcuatitán | 1 577 | 18.1908333 | -92.9611111 |
| 36 | Tepetitán | 1 543 | 17.8188889 | -92.3725 |

**Tabla 5 (continuación).** Resultados de población afectada por localidad para Tabasco.

| | Nombre | Habitantes | Latitud | Longitud |
|---|---|---|---|---|
| 37 | Tecoluta 2a. Sección | 1 517 | 18.25 | -93.0194444 |
| 38 | Tránsito Tular | 1 465 | 18.3402778 | -93.3975 |
| 39 | Gobernador Cruz | 1 454 | 18.435 | -92.8716667 |
| 40 | Nuevo Torno Largo | 1 381 | 18.4327778 | -93.1633333 |
| 41 | Quintín Aráuz | 1 353 | 18.3236111 | -92.5661111 |
| 42 | Nueva División del Bayo (Guatemala) | 1 333 | 17.8402778 | -92.4888889 |
| 43 | Chichicastle 1a. Sección | 1 318 | 18.3094444 | -92.4480556 |
| 44 | Acachapan y Colmena 3a. Sección | 1 239 | 18.0477778 | -92.7780556 |
| 45 | Chiltepec (Sección Banco) | 1 214 | 18.4275 | -93.1130556 |
| 46 | Occidente (San Francisco) | 1 199 | 18.3313889 | -93.2522222 |
| 47 | Pénjamo | 1 196 | 18.4347222 | -93.0930556 |
| 48 | Libertad de Allende | 1 173 | 18.4086111 | -92.8269444 |
| 49 | Estancia | 1 170 | 18.1772222 | -92.8127778 |
| 50 | Oriente (San Cayetano) | 1 128 | 18.3480556 | -93.2052778 |
| 51 | Reforma 2a. Sección (Santa María) | 1 128 | 18.3352778 | -93.0461111 |
| 52 | Las Flores 3a. Sección (El Cerro) | 1 099 | 18.425 | -93.2797222 |
| 53 | José María Pino Suárez 2a. Sección | 1 098 | 18.3838889 | -93.3794444 |
| 54 | Chiltepec (Sección Tanque) | 1 077 | 18.4283333 | -93.0913889 |
| 55 | Unión y Libertad | 1 069 | 17.9447222 | -92.6416667 |
| 56 | Potreritos | 1 068 | 18.3113889 | -93.2819444 |
| 57 | La Victoria | 1 044 | 18.5905556 | -92.6341667 |
| 58 | El Escribano | 1 034 | 18.4105556 | -93.2216667 |
| 59 | Jalapita | 1 033 | 18.4133333 | -92.9938889 |
| 60 | La Sábana | 1 004 | 18.4416667 | -92.8877778 |
| 61 | Ignacio Zaragoza 1a. Sección | 1 000 | 18.3316667 | -93.3355556 |
| | **Total** | 199,491 | | |
| | **%Pob.** | 10.02% | | |

tes. El estado con mayor superficie afectada es Tabasco con más del 21% de su territorio, mientras que el que tendrá una población más vulnerable será Veracruz, con más de 1 millón de habitantes si el escenario de incremento en 1 m se presentará hoy en día. Quintana Roo, por su parte, tendrá proporcionalmente un mayor impacto pues sufrirá los efectos en el 81.1% de su población. En el caso de Tabasco y Campeche, la línea costera sufrirá un importante retroceso, lo que impactará también en las poblaciones que en la actualidad no están en la costa, además de que se producirá una importante contaminación de las fuentes de agua dulce y la salinización de los suelos.

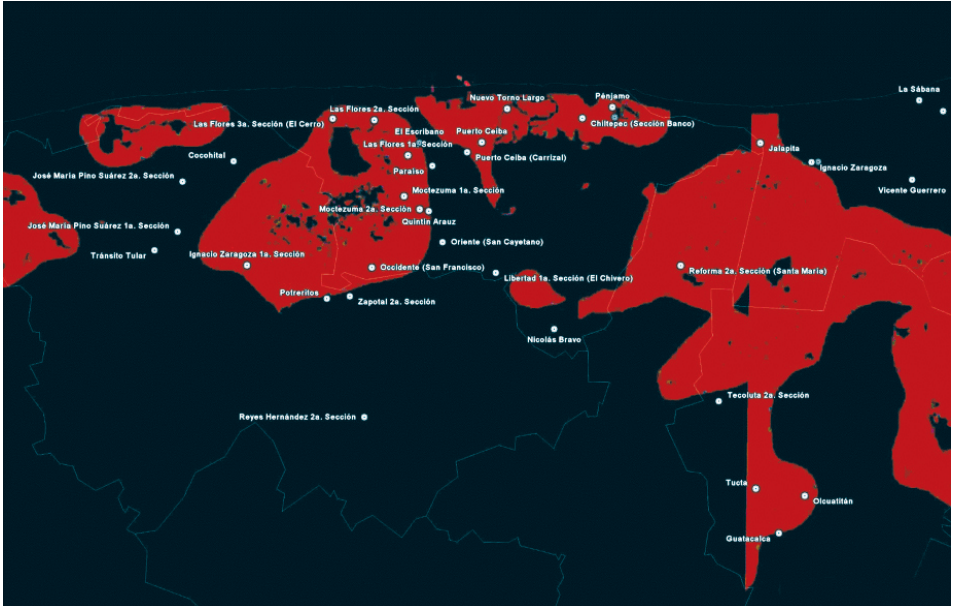

**Figura 8.** Tabasco. Poblaciones costeras de Tabasco afectadas. Nótese la gran dispersión de los asentamientos así como la afectación sobre asentamientos lejanos a la costa.

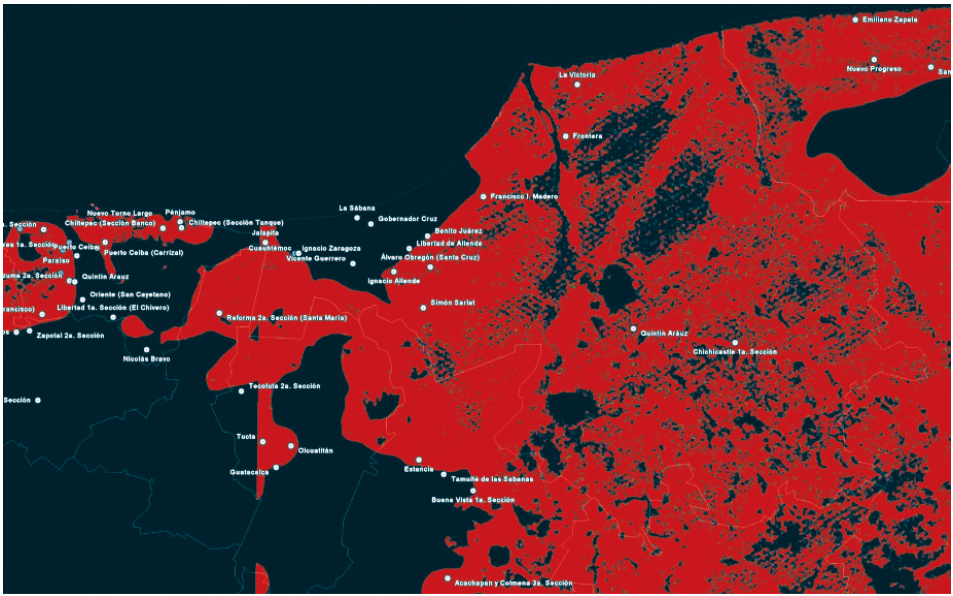

**Figura 9.** Tabasco- Campeche. Se observa una enorme afectación debido a que predominan las zonas bajas. Nótese la afectación sobre asentamientos distantes a la costa.

**Tabla 6.** Resultados de población afectada por localidad para Campeche.

|   | Nombre | Habitantes | Latitud | Longitud |
|---|---|---|---|---|
| 1 | Campeche | 211 671 | 19.8422222 | -90.5316667 |
| 2 | Ciudad del Carmen | 154 197 | 18.6433333 | -91.8308333 |
| 3 | Champotón | 27 235 | 19.3555556 | -90.7233333 |
| 4 | Seybaplaya | 8 285 | 19.6383333 | -90.6877778 |
| 5 | Sabancuy | 6 159 | 18.9738889 | -91.1794444 |
| 6 | Isla Aguada | 4 688 | 18.7847222 | -91.4916667 |
| 7 | Nuevo Progreso | 4 492 | 18.6216667 | -92.2888889 |
| 8 | Villa Madero | 3 507 | 19.5266667 | -90.7013889 |
| 9 | San Antonio Cárdenas | 3 319 | 18.6141667 | -92.2225 |
| 10 | Palizada | 3 061 | 18.2555556 | -92.0916667 |
| 11 | Sihochac | 2 631 | 19.5013889 | -90.5861111 |
| 12 | Ley Federal de Reforma Agraria | 2 398 | 19.0622222 | -90.8080556 |
| 13 | Atasta | 2 096 | 18.6197222 | -92.1041667 |
| 14 | Checubul | 1 541 | 18.8233333 | -91.0116667 |
| 15 | Francisco Villa (Mamantel) | 1 208 | 18.5244444 | -91.09 |
| 16 | El Aguacatal (Chumpán) | 1 189 | 18.2138889 | -91.5105556 |
| 17 | Emiliano Zapata | 1 126 | 18.6652778 | -92.3111111 |
| 18 | CERESO San Francisco Kobén | 1 072 | 19.9111111 | -90.4213889 |
| 19 | Imí | 1 035 | 19.8722222 | -90.4711111 |
|   | **Total** | 440 910 |   |   |
|   | **%Pob.** | 58.41% |   |   |

## Recomendaciones

Se recomienda realizar estudios mas detallados para conocer la topografía del terreno aledaño a los núcleos de población importante con mejor precisión usando otras tecnologías como el lidar[5]. Lo anterior es necesario para conocer a detalle las zonas y sus niveles de impacto así como las zonas que podrían servir para la instalación de nuevas zonas urbanas. En este caso, es importante que la reubicación sea lo mejor planeada y organizada posible con el fin de optimizar recursos y que se pueda resolver también el problema de la dispersión poblacional en las regiones rurales [6].

---

[5] Hinkel J., y R.J. Klein, 2009. Integrating knowledge to assess coastal vulnerability to sea-level rise: The development of the diva tool. *Global Environmental Change*, 19: 384-395.
[6] Bosello, F., R. Roson, y R. Tol, 2007. Economy-wide estimates of the implications of climate change: sea level rise. *Environmental and Resource Economics*, 37: 549-571.

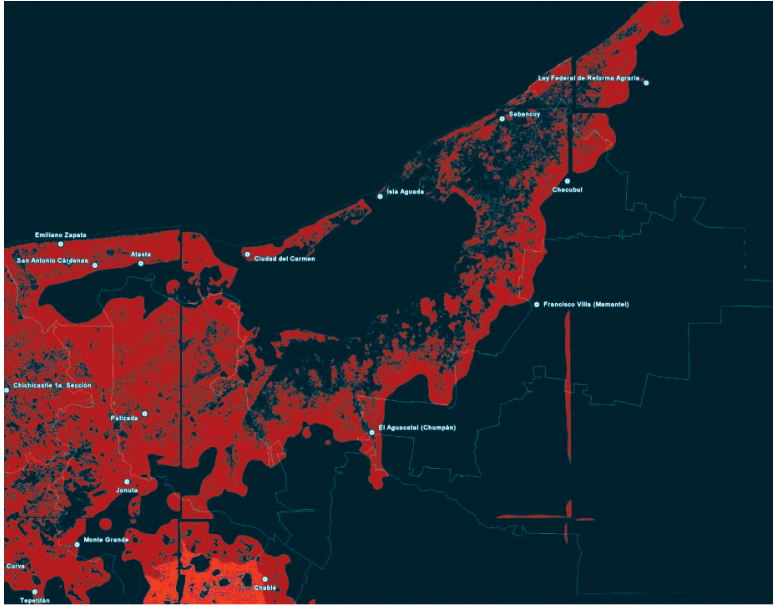

**Figura 10.** Campeche. Se observa la afectación de Ciudad del Carmen y Sabancuy.

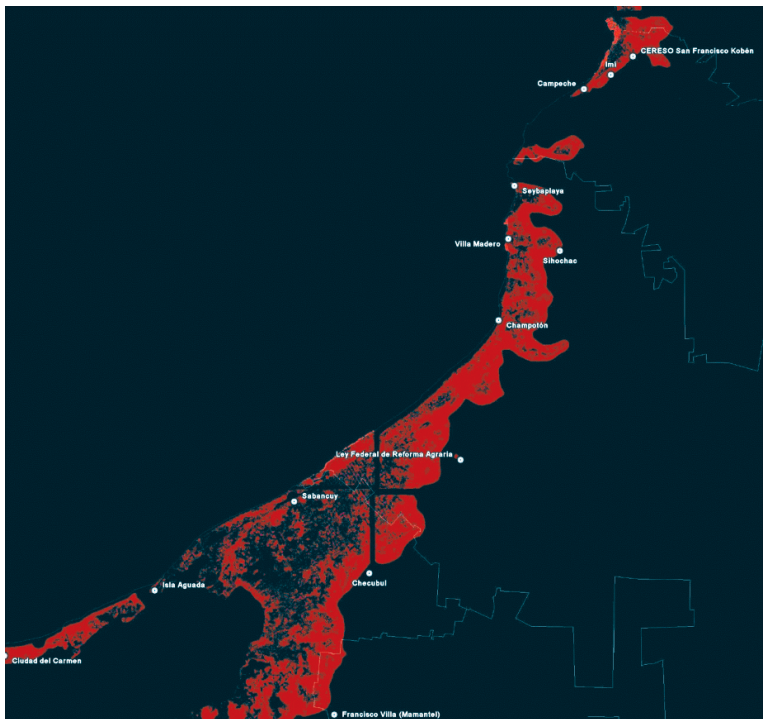

**Figura 11.** Campeche. Nótese la afectación sobre Campeche y Champotón.

Tabla 7. Resultados de población afectada por localidad para Yucatán.

|    | Nombre                      | Habitantes | Latitud    | Longitud    |
|----|-----------------------------|------------|------------|-------------|
| 1  | Progreso                    | 35 519     | 21.2827778 | -89.6636111 |
| 2  | Chicxulub (Chicxulub Puerto)| 5 052      | 21.2938889 | -89.6083333 |
| 3  | Campestre Flamboyanes       | 3 022      | 21.21      | -89.6577778 |
| 4  | Chelem                      | 3 017      | 21.2688889 | -89.7430556 |
| 5  | Dzilam de Bravo             | 2 188      | 21.3925    | -88.8913889 |
| 6  | Río Lagartos                | 2 127      | 21.5975    | -88.1577778 |
| 7  | San Felipe                  | 1 769      | 21.5672222 | -88.2311111 |
| 8  | El Cuyo                     | 1 748      | 21.5158333 | -87.6783333 |
| 9  | Chuburná                    | 1 720      | 21.2533333 | -89.8166667 |
| 10 | Sisal                       | 1 672      | 21.1652778 | -90.0305556 |
| 11 | Telchac Puerto              | 1 618      | 21.3405556 | -89.2630556 |
| 12 | Las Coloradas               | 1 068      | 21.6083333 | -87.9897222 |
| 13 | Celestún                    | 6 243      | 20.8591667 | -90.4        |
|    | **Total**                   | 66 763     |            |             |
|    | **%Pob.**                   | 3.67%      |            |             |

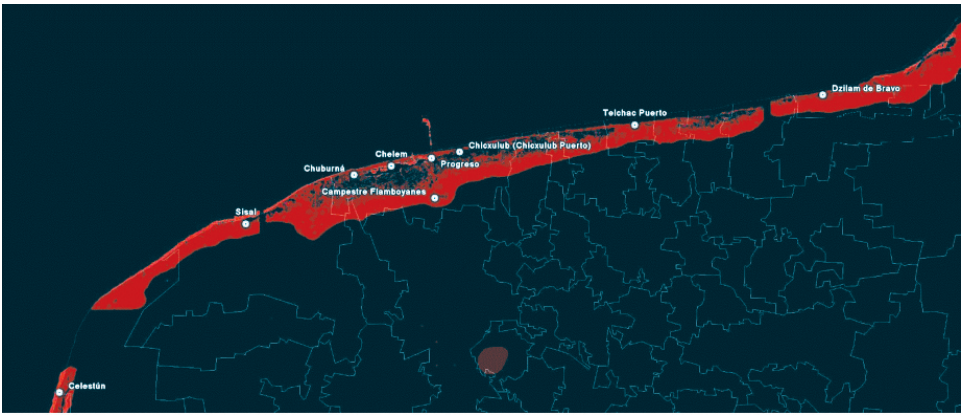

**Figura 12.** Yucatán. En este caso, se observa la afectación de Progreso y Celestún.

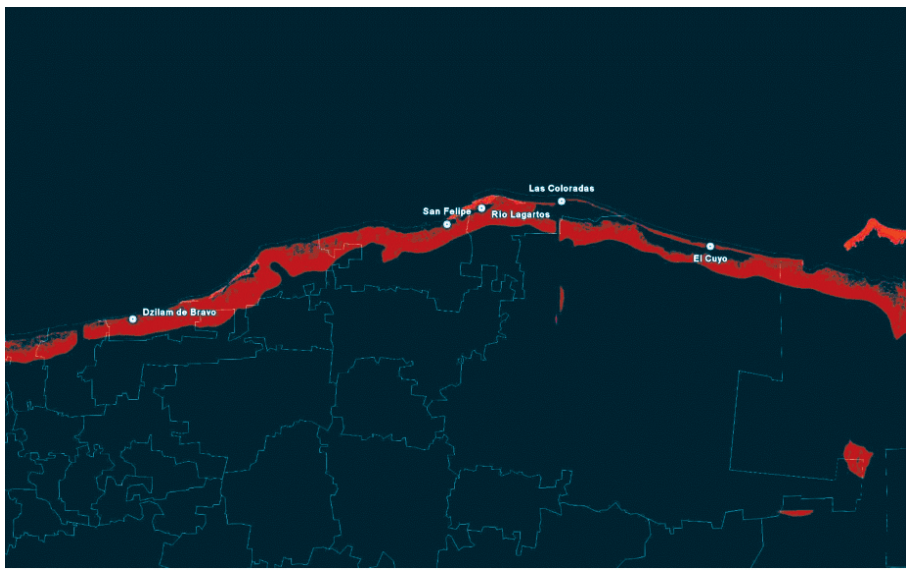

**Figura 13.** Yucatán. Se observa la afectación a todo lo largo de la costa aunque el avance de la línea costera es relativamente pequeño.

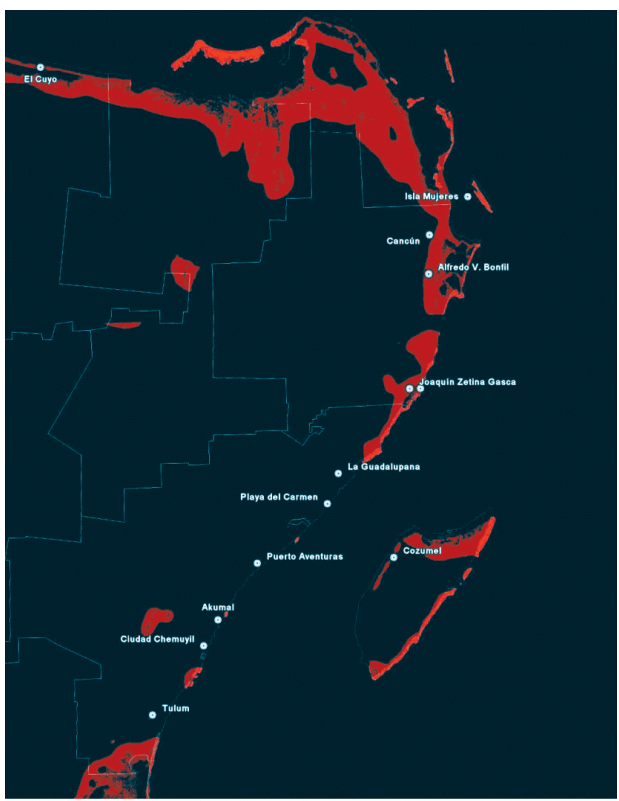

**Figura 14.** Quintana Roo. La mayoría de los municipios y las poblaciones más importantes son afectadas.

| | **Tabla 8.** Resultados de población afectada por localidad para Quintana Roo. | | | |
|---|---|---|---|---|
| | **Nombre** | Habitantes | Latitud | Longitud |
| 1 | Cancún | 526 701 | 21.1605556 | -86.8475 |
| 2 | Chetumal | 136 825 | 18.5036111 | -88.3052778 |
| 3 | Playa del Carmen | 100 383 | 20.6275 | -87.0811111 |
| 4 | Cozumel | 71 401 | 20.5166667 | -86.9416667 |
| 5 | Tulum | 14 790 | 20.2119444 | -87.4658333 |
| 6 | Alfredo V. Bonfil | 13 822 | 21.0825 | -86.8513889 |
| 7 | Isla Mujeres | 11 147 | 21.2355556 | -86.7627778 |
| 8 | Bacalar | 9 833 | 18.6769444 | -88.3952778 |
| 9 | Joaquín Zetina Gasca | 6 629 | 20.8536111 | -86.8986111 |
| 10 | La Guadalupana | 5 892 | 20.6875 | -87.0561111 |
| 11 | Calderitas | 4 446 | 18.5544444 | -88.2583333 |
| 12 | Limones | 1 961 | 19.0241667 | -88.1083333 |
| 13 | Cacao | 1 915 | 18.1927778 | -88.695 |
| 14 | Subteniente López | 1 890 | 18.4936111 | -88.3930556 |
| 15 | Xul-Ha | 1 838 | 18.5516667 | -88.4638889 |
| 16 | Carlos A. Madrazo | 1 769 | 18.5022222 | -88.5225 |
| 17 | Pucté | 1 757 | 18.2333333 | -88.6613889 |
| 18 | Puerto Aventuras | 1 629 | 20.5116667 | -87.2341667 |
| 19 | Ucum | 1 345 | 18.5030556 | -88.5183333 |
| 20 | Sabidos | 1 265 | 18.3558333 | -88.5894444 |
| 21 | Ciudad Chemuyil | 1 239 | 20.3486111 | -87.3530556 |
| 22 | Akumal | 1 198 | 20.4 | -87.3211111 |
| 23 | Puerto Morelos | 1 097 | 20.8536111 | -86.8752778 |
| | **Total** | **920 772** | | |
| | **%Pob** | **81.10%** | | |

Se sugiere impulsar estudios orientados de forma que se comprenda los efectos locales del incremento del nivel del mar por mareas, corrientes marinas y procesos de erosión de la franja costera que permita tener una visión clara de la dinámica costera de cada lugar, principalmente, las más vulnerables[7].

Es primordial el diseño de planes de desarrollo urbano que tomen en cuenta las zonas vulnerables aquí reportadas, de lo contrario, un número mayor de habitantes podrian verse afectados. De hecho se deberían considerar políticas que prohíban los asentamientos humanos en dichas áreas.

---

[7] McGranahan, G., D. Balk, y B. Anderson, 2007. The rising tide: assessing the risks of climate change and human settlements in low elevation coastal zones. *Environment and Urbanization*, 19: 17-37.

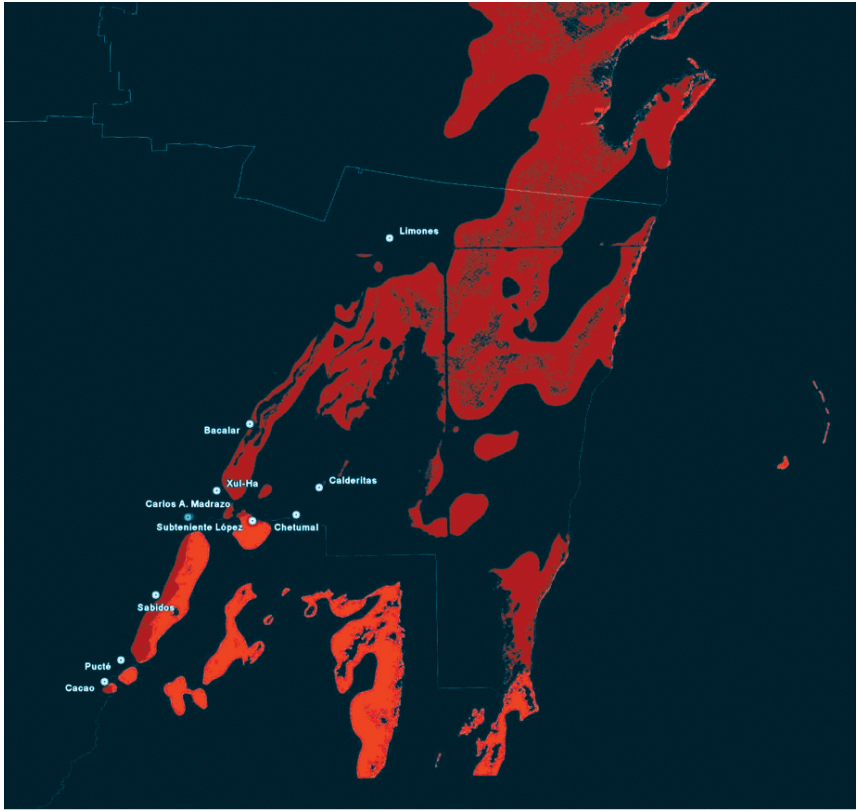

**Figura 15.** Quintana Roo. Se observa una gran vulnerabilidad alrededor de Chetumal.

Se debe hacer un inventario de infraestructura vulnerable y así realizar una mejor evaluación de los impactos y conocer que áreas se pueden proteger con defensas costeras y cuales son más rentables de dejar sin protección. De esta forma, los tomadores de decisiones tendran información confiable que les permita decidir las estrategias a seguir, en términos de costo y seguridad de la población[12].

Es muy importante que con la información obtenida por estudios como el presente y posteriores, se desarrolle una sistema de información geográfica (SIG) dedicado a los problemas de cambio climático, de acceso público, para facilitar la educación de la población y la toma de decisiones.

---

[8] Cayan, D., P. Bromirski, K. Hayhoe, M. Tyree, M. Dettinger, y R. Flick, 2008. Climate change projections of sea level extremes along the California coast. *Climatic Change*, 87: 57-73.

# Apéndice I. Lista de archivos hgt utilizados

| | |
|---|---|
| N17W092.hgt | N20W087.hgt |
| N17W093.hgt | N20W088.hgt |
| N17W094.hgt | N20W090.hgt |
| N17W095.hgt | N20W091.hgt |
| N17W096.hgt | N20W097.hgt |
| N18W088.hgt | N20W098.hgt |
| N18W089.hgt | N21W087.hgt |
| N18W091.hgt | N21W088.hgt |
| N18W092.hgt | N21W089.hgt |
| N18W093.hgt | N21W090.hgt |
| N18W094.hgt | N21W091.hgt |
| N18W095.hgt | N21W098.hgt |
| N18W096.hgt | N21W099.hgt |
| N18W097.hgt | N22W098.hgt |
| N19W088.hgt | N22W099.hgt |
| N19W091.hgt | N23W098.hgt |
| N19W092.hgt | N23W099.hgt |
| N19W096.hgt | N24W098.hgt |
| N19W097.hgt | N24W099.hgt |
| N19W098.hgt | N25W098.hgt |